\documentclass[twocolumn,showpacs,preprintnumbers,amsmath,amssymb]{revtex4}

\usepackage{graphicx}

\begin{document}


\title{Zero-range model of traffic flow}

\author{J. Kaupu\v{z}s}
 \email{kaupuzs@latnet.lv}
\affiliation{Institute of Mathematics and Computer Science, University of Latvia,\\
LV--1459 Riga, Latvia
}
\author{R. Mahnke}
 \email{reinhard.mahnke@uni-rostock.de}
\affiliation{Institut f\"ur Physik, Universit\"at Rostock, D--18051 Rostock, Germany
}

\author{R. J. Harris}
 \email{r.harris@fz-juelich.de}
\affiliation{Institut f\"ur Festk\"orperforschung, Forschungszentrum J\"ulich,\\ 
D--52425 J\"ulich, Germany
}%

\date{\today}

\begin{abstract}
A multi--cluster model of traffic flow is studied, in which the motion
of cars is described by a stochastic master equation. Assuming that the
escape rate from a cluster depends only on the cluster size, the
dynamics of the model is directly mapped to the mathematically
well-studied zero-range process. Knowledge of the asymptotic behaviour
of the transition rates for large clusters allows us to apply an
established criterion for phase separation in one-dimensional driven
systems. 
The distribution over cluster sizes in our zero-range model
is given by a one--step master equation in one dimension. It
provides an approximate mean--field dynamics, which,
however, leads to the exact stationary state. Based on
this equation, we have calculated the critical density at which phase
separation takes place.  We have shown that within a
certain range of densities above the critical value a metastable
homogeneous state exists before coarsening sets in. Within this
approach we have estimated the critical cluster size and the mean
nucleation time for a condensate in a large system. The metastablity in
the zero-range process is reflected in a metastable branch of the
fundamental flux--density diagram of traffic flow. Our work thus
provides a possible analytical description of traffic jam formation as
well as important insight into condensation in the zero-range process.
\end{abstract}

\maketitle

\section{Introduction}

The development of traffic jams in vehicular flow is an everyday example
of the  occurence of phase separation in low-dimensional driven systems,
a topic which  has attracted much recent interest (see,
e.~g.,~\cite{Mukamel00,Schutz03} and references  therein). 
In~\cite{S1}
the existence of phase separation is related to the  size-dependence of
domain currents and a quantitative criterion is obtained by  considering
the zero-range process (ZRP) as a generic model for domain dynamics.  
Phase separation corresponds to the phenomenon of condensation in the
ZRP  (see~\cite{Evans05} for a recent review) whereby a macroscopic
proportion of particles  accumulate on a single site.

In this paper we use such a zero-range picture to study the phase
separation in traffic flow.   A stochastic master equation approach in
the spirit of the probabilistic  description of
transportation~\cite{physrep} enables us to  calculate the critical
parameters of our model.  We pay particular attention to the initial 
nucleation of a condensate in the ZRP, complementing previous
work~\cite{S3,Godreche03} on  the late-stage cluster coarsening. 
Significantly, we find that prior to condensation the  system can exist
in a homogeneous metastable state and we provide estimates of the
critical  cluster size and mean nucleation time.
 Finally, we apply
these results to the description of traffic, obtaining a fundamental 
flux-density diagram which includes a metastable branch.
 Metastability
and hysteresis effects have been observed in real traffic,  see,
e.~g.,~\cite{cssrev,Helbing01} for discussion of empirical data and the
various different  modelling approaches.  For previous work focusing on
the description of jam formation as a  nucleation process,
see~\cite{mk99,mkf03}.

\section{The model}

We consider a model of traffic flow, where cars are moving along
a circular road. Each car occupies a certain length of road
$\ell$. We divide the whole road of total length $L$ in to
cells of size $\ell$.
Each cell can be either empty or occupied by a car, just as in cellular 
automaton traffic
models (see, e.~g.,~\cite{cssrev,Helbing01} and references therein).  
Most such models use a discrete-time update rule, for example, see~\cite{Levine04} 
for a class of traffic models related to a parallel updating version of the ZRP.  
In contrast, we consider the development of our system in continuous time.
The probability per unit time for each car to move to the next cell
is given by a
certain transition rate, which depends on the actual
configuration of empty and occupied cells. 
This configuration is characterised
by the cluster distribution. An uninterrupted string of $n$ occupied cells,
bounded by two empty cells, is called a cluster of size $n$. The clusters
of size $n=1$ are associated with freely moving cars.
The first car in each cluster is allowed to move forward by one cell. 
The transition rate $w_n$ of this stochastic event depends on the
size $n$ of the cluster to which the car belongs. In this case $w_1$
is the mean of the inverse time necessary for a free car to move
forward by one cell.
The transition rate $w_1$ is related to the distribution of velocities
in the free flow regime or phase, which is characterised by a certain
car density $c_{\mbox{\scriptsize free}}$.
For small densities, expected in the free flow 
phase in real traffic, the interaction between cars is weak and, therefore, 
the transition rate $w_1$ depends only weakly on the 
density $c_{\mbox{\scriptsize free}}$.
Hence, in the first approximation we may assume that $w_1$ is a 
constant.

This model can be directly mapped to the zero-range process (ZRP).
Each vacancy (empty cell) in the original model is related to a box
in the zero-range model. The number of boxes is fixed, and each box
can contain an arbitrary number of particles (cars), which is equal
to the size of the cluster located to the left 
(if cars are moving to the right) of
the corresponding vacancy in the original model.
If this vacancy has another vacancy to the left, then it means that 
the box is empty. Since the boundary conditions are periodic in the original
model, they remain periodic also in the zero-range model. 
In this representation, one particle moves from a given box
to the right with transition rate $w_n$, which
depends only on the number of particles $n$ in this box.
In the grand canonical ensemble, where
the total number of particles is allowed to fluctuate, 
the stationary 
distribution
over the cluster--size configurations is the product of independent 
distributions for individual boxes. The probability that there
are just $n$ particles in a box in a homogeneous phase is
\cite{Spitzer70,Evans00}
$P(n) \propto z^n / \prod_{m=1}^n w_m$ for $n>0$, $P(0)$ being given
by the normalisation condition. Here $z=e^{\mu/k_BT}$ is the fugacity -- a parameter
which controls the mean number of particles in the system.

\section{Master equation}
\label{sec:master}

This result can be obtained and interpreted
within the stochastic master equation approach~\cite{Evans05}.
Assuming the statistical independence of the distributions in 
different boxes, we have a multiplicative ansatz
\begin{equation}
P_2(k,m,t) = P(k,t) \, P(m,t)
\label{eq:mult} 
\end{equation}
for the joint probability $P_2(k,m,t)$
that there are $m$ particles in one box and $k$ particles
in the neighbouring box on the left at time $t$.
This approximation leads to the mean--field dynamics described 
by the master equations~\cite{Evans05}
\begin{eqnarray}
\label{eq:master}
\frac{\partial P(n,t)}{\partial t} &=&  \langle w \rangle P(n-1,t) 
+ w_{n+1} P(n+1,t) \nonumber \\
&-&  [\langle w \rangle + w_n] P(n,t) \quad : \quad n\ge 1 \;, \\
\label{eq:master0}
\frac{\partial P(0,t)}{\partial t} &=&  
w_{1} P(1,t) - \langle w \rangle P(0,t)  \;, 
\end{eqnarray} 
where 
\begin{equation}
\langle w \rangle (t) = \sum\limits_{k=1}^{\infty} w_k P(k,t)
\label{eq:w}
\end{equation}
is the mean inflow rate in a box.
The ansatz~(\ref{eq:mult}) is an exact property of the stationary 
state of the grand canonical ensemble or, alternatively, of an 
infinitely large system
\cite{Spitzer70}.  Hence, in these cases, the master equations~(\ref{eq:master}) 
and~(\ref{eq:master0}) give the exact stationary state while providing
a mean-field approximation to the dynamics of reaching it.

The stationary solution $P(n)$ corresponding to $\partial P(n,t)/ \partial t = 0$
can be found recursively, starting from $n=0$. It yields the known result 
\cite{Spitzer70,Evans00,Evans05}
\begin{equation}
P(n) = P(0) \, \langle w \rangle^n \prod\limits_{m=1}^n \frac{1}{w_m} 
\label{eq:Pk}
\end{equation}
for $n >0$, where $P(0)$ is found from the normalisation condition.

Denoting by $M$ the number of boxes, which corresponds to the
number of vacancies in the original model, the mean number of cars
on the road is given by $\langle N \rangle = M \langle n \rangle$,
where 
\begin{equation}
\langle n \rangle = \sum\limits_{n=1}^{\infty} n \, P(n)
\label{eq:kav}
\end{equation}
is the average number of particles in a box.
Note that in the grand canonical ensemble the total number of cars
as well as the length of the road $L$ fluctuate. For the mean value,
measured in units of $\ell$,
we have $\langle L \rangle = M + \langle N \rangle$. Hence, the average
density of cars is 
\begin{equation}
c = \frac{\langle N \rangle}{\langle L \rangle} 
= \frac{\langle n \rangle}{1 + \langle n \rangle} \;.
\label{eq:c}
\end{equation} 
According to~(\ref{eq:c}), (\ref{eq:kav}), and~(\ref{eq:Pk}),
we have the following relation
\begin{equation}
\frac{c}{1-c} = \frac{\sum\limits_{n=1}^{\infty}
n \langle w \rangle^n \prod\limits_{m=1}^n \displaystyle{\frac{1}{w_m}} }
{1 + \sum\limits_{n=1}^{\infty} 
\langle w \rangle^n \prod\limits_{m=1}^n \displaystyle{\frac{1}{w_m}} }
\label{eq:uuu}
\end{equation}
from which the stationary mean inflow rate $\langle w \rangle$ can be
calculated at a given average density $c$.

\section{Transition rates and phase separation}
\label{sec:trr}

Now we make the following choice for the transition rate dependence
on the cluster size $n$:
\begin{equation}
w_n = w_{\infty} \left( 1 + \frac{b}{n^{\sigma}} \right)
\hspace{3ex} \mbox{for} \hspace{2ex} n \ge 2 \;,
\label{eq:rates}
\end{equation}
the value of $w_1$ being given separately, since it is related
to the motion of uncongested cars, whereas $w_n$ with $n \ge 2$
represents the
escaping from a jam of size $n$.  Although an individual driver does
not know how many cars are jammed behind him, the effective current of
cars from a jam, represented by $w_n$, is a collective effect which is
expected to depend on the correlations and internal struture (e.~g.,
distribution of headways) within the cluster~\cite{S1}. A monotonously
decreasing dependence on cluster size, such as~\eqref{eq:rates}, can be
considered as a type of slow-to-start rule---the longer a car has been
stationary the larger the probability of a delay when starting
(cf.~\cite{Takayasu93,Benjamin96,Barl98,LEC98}).

We now explore the consequences of the choice~(\ref{eq:rates}) in terms 
of the ZRP phase behaviour and its implications for the description of traffic 
flow. In numerical calculations we have assumed 
$w_{\infty}=1/\tau_{\infty}=1$ and $w_1=5$, by choosing the time constant
$\tau_{\infty}$ as a time unit, 
whereas the control parameters $b$ and $\sigma$ have been varied.

\begin{figure}
\begin{center}
\includegraphics[scale=0.3]{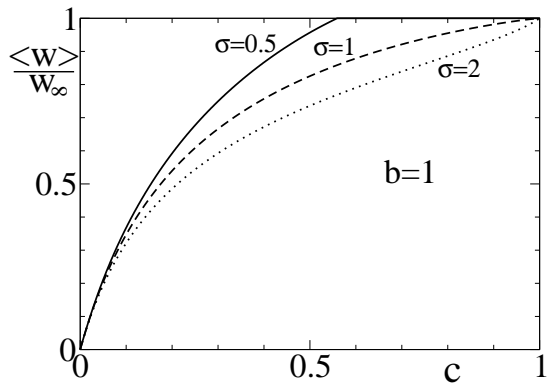}
\end{center}
\caption{ 
$\langle w \rangle / w_{\infty}$ vs density $c$ at $b=1$ for different $\sigma$.}
\label{w_1}
\end{figure}

If $\sigma >1$, as well as for $b \le 2$ at $\sigma =1$, 
Eq.~(\ref{eq:uuu}) has a solution for any density $0<c<1$
(see dashed and dotted curves in Fig.~\ref{w_1}).
This implies that the
homogeneous phase is stable in the whole range of densities,
i.~e., there is no phase transition in a strict sense.
If $\sigma <1$ (solid curve in Fig.~\ref{w_1}), as well as for $b >2$ at $\sigma =1$,
$\langle w \rangle / w_{\infty}$ reaches $1$ at a critical
density $0<c_{cr}<1$, and there is no physical solution
of~(\ref{eq:uuu}) for $c>c_{cr}$. 
This means that the homogeneous phase cannot accommodate a
larger density of particles and condensation takes place at $c>c_{cr}$. 

This behaviour underlies the known criterion for phase 
separation in one--dimensional driven systems~\cite{S1}. For illustration,
we comment that in the multi--cluster model considered in~\cite{km00}
the transition rates do not depend on the cluster sizes, only
the inflow rate in a cluster depends on the overall car density
and fraction of congested cars. This corresponds to the case $b=0$,
where, according to the criterion, no macroscopic phase separation takes 
place -- in agreement
with the theoretical conclusions and simulation results of~\cite{km00}. 
In contrast, a class of microscopic models was introduced in~\cite{S2} 
where correlations within the domain (jam) give rise to currents of 
the form~\eqref{eq:rates} with $\sigma=1$ and $b>2$; 
phase separation is then observed.

At $c<c_{cr}$ in our model the cluster distribution function $P(n)$
decays exponentially fast for large $n$  
whereas the decay is slower at $c=c_{cr}$.
It is
well known that the decay in this case is power--like for 
 $\sigma=1$,
i.~e., $P(n) \sim n^{-b}$~\cite{S1}. For $0 < \sigma < 1$, we find
that the leading behaviour is stretched exponential, i.~e., 
 $P(n) \sim
\exp \left[ -b \, n^{1-\sigma}/(1-\sigma) \right]$, in agreement with
the result stated in~\cite{Evans05}.  Within our approach we have also
derived the sub-leading terms which turn out to be relevant for $0 <
\sigma < 1/2$.  This calculation is presented in the Appendix and
further illustrates the rich behaviour of the model as $\sigma$ is
varied.
%
%

At $\langle w \rangle / w_{\infty}=1$
the inflow $\langle w \rangle$ in a macroscopic cluster of size $n \to \infty$ 
is balanced with the outflow $w_{\infty}$. 
This means that at overall density $c>c_{cr}$ the
homogeneous phase with density $c_{cr}$ is in equilibrium with
a macroscopic cluster, represented by one of the boxes containing
a non--vanishing fraction of all particles in the thermodynamic limit
\cite{S3,Jeon00}.
Hence, $\langle w \rangle / w_{\infty}=1$ holds in the phase
coexistence regime at $c>c_{cr}$. 

\begin{figure}
\begin{center}
\includegraphics[scale=0.3]{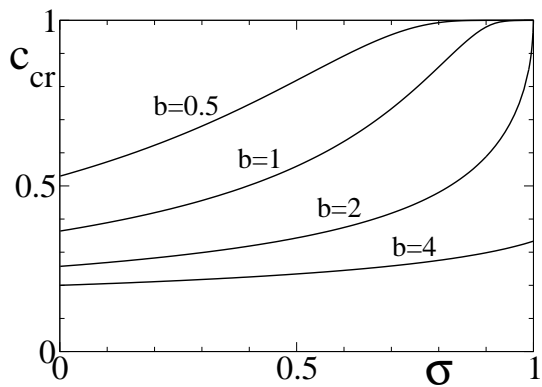}
\end{center}
\caption{Critical density as a function of control parameter $\sigma$
for different values of $b$.}
\label{ccr_sig}
\end{figure}

\begin{figure}
\begin{center}
\includegraphics[scale=0.3]{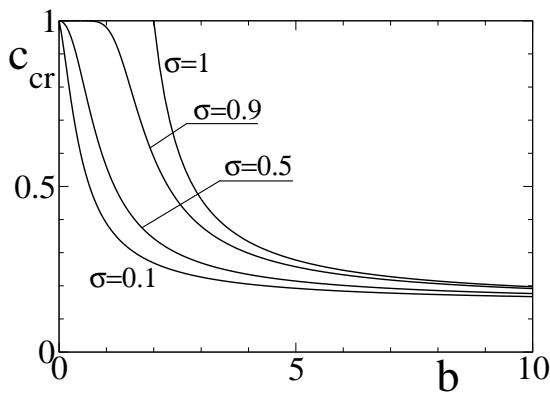}
\end{center}
\caption{Critical density as a function of control parameter $b$
for different values of $\sigma$.}
\label{ccr_b}
\end{figure}

According to~(\ref{eq:c}), the critical density $c_{cr}$ is given by
\begin{equation}
c_{cr} = \frac{\langle n \rangle_{cr}}{1 + \langle n \rangle_{cr}} \;,
\label{eq:ccr}
\end{equation}
where $\langle n \rangle_{cr}$ is the mean cluster size at the critical
density. Since $\langle w \rangle = w_{\infty}$ holds in this case, we 
have
\begin{equation}
\langle n \rangle_{cr} = \frac{\sum\limits_{n=1}^{\infty}
n \, w_{\infty}^n \prod\limits_{m=1}^n \displaystyle{\frac{1}{w_m}} }
{1 + \sum\limits_{n=1}^{\infty} 
w_{\infty}^n \prod\limits_{m=1}^n \displaystyle{\frac{1}{w_m}} } \;.
\label{eq:kcr}
\end{equation}
The critical density, calculated numerically
from~(\ref{eq:ccr}) and~(\ref{eq:kcr})
as a function of parameters $\sigma$ and $b$, is shown in 
Figs.~\ref{ccr_sig} and~\ref{ccr_b}, respectively. 
In contrast to the situations discussed previously in the literature, in
our model $w_1$ is not given by the general formula~(\ref{eq:rates}) but
is an independent parameter. This distinction leads to quantitatively
different results, for example, the critical density for $\sigma=1$ is
analytically shown to be~\cite{GSprivate}
\begin{equation}
c_{cr}=\frac{b(b+1)}{(b-1)[2(b+1)+w_1 (b-2)]} \; .
\end{equation}

\section{Metastability}

Suppose that at the initial time moment $t=0$ the system is in a
homogeneous state with overall density slightly larger
than $c_{cr}$. Here we study the development of such a state in the 
mean--field approximation provided by~(\ref{eq:master}) 
and~(\ref{eq:master0}).
With this initial condition, the mean inflow rate in a box 
$\langle w \rangle$ is slightly larger than 
that at $c=c_{cr}$, i.~e., $\langle w \rangle=w_{\infty}+\varepsilon$
holds with small and positive $\varepsilon$.
Hence, only large clusters with $w_n < w_{\infty}+\varepsilon$
have a stable tendency to grow,
whereas any smaller cluster typically (except a rare case) fluctuates 
until it finally dissolves. In other words, the initially
homogeneous system with no large clusters can stay in this metastable
supersaturated state for a long time until a large stable
cluster appears due to a rare fluctuation.

Neglecting the fluctuations, the time development of the size
$n$ of a cluster is described by the deterministic equation
\begin{equation}
\frac{dn}{dt} = \langle w \rangle - w_n \;.
\label{eq:det}
\end{equation}
According to this equation, the undercritical clusters
with $n<n_{cr}$ tend to dissolve, whereas the overcritical
ones with $n>n_{cr}$ tend to grow, where 
the critical cluster size $n_{cr}$ is given by the condition
\begin{equation}
\langle w \rangle = w_{n_{cr}} \;.
\label{eq:kcrcond}
\end{equation}
Using~(\ref{eq:rates}) yields
\begin{equation}
n_{cr} \simeq \left( \frac{b}{\langle w \rangle / w_{\infty} -1} \right)^{1/\sigma} \;.
\label{eq:k_cr}
\end{equation} 
In this case $n_{cr}$ is rounded to an integer value.

\begin{figure}
\begin{center}
\includegraphics[scale=0.3]{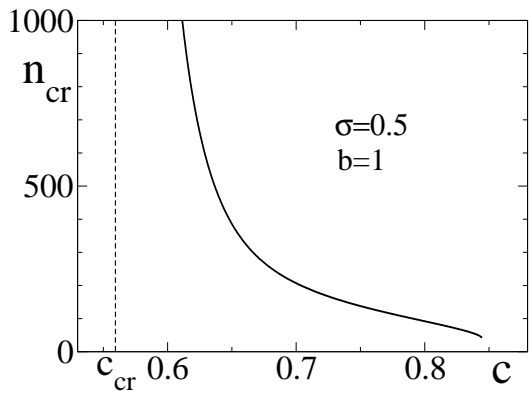}
\end{center}
\caption{Critical cluster size vs density at $\sigma=0.5$ and $b=1$.
The critical density is indicated by a vertical dashed line.}
\label{kcr_sig}
\end{figure}

\begin{figure}
\begin{center}
\includegraphics[scale=0.3]{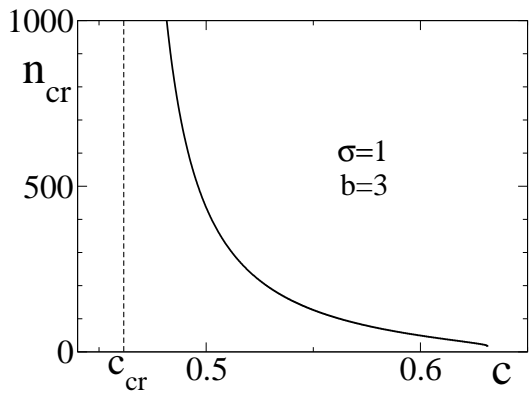}
\end{center}
\caption{Critical cluster size vs density at $\sigma=1$ and $b=3$.
The critical density is indicated by a vertical dashed line.}
\label{kcr_b}
\end{figure}

This deterministic approach describes only the most probable scenario 
for an arbitrarily choosen cluster of a given size. It 
does not allow one to obtain the distribution over cluster sizes:  
the deterministic equation~(\ref{eq:det}) suggests that all clusters shrink 
to zero size if they are smaller than $n_{cr}$ at the beginning, whereas 
the real size distribution arises from the competition between opposite stochastic 
events of shrinking and growing.
Assuming that the distribution of relatively small clusters 
contributing to $\langle n \rangle$
is quasi--stationary, i.~e., that the detailed balance 
(equality of the terms in~(\ref{eq:master}) 
and~(\ref{eq:master0}) describing opposite stochastic events) 
for these clusters is almost reached before any cluster with $n > n_{cr}$ 
has appeared, we have
\begin{equation}
\langle n \rangle \simeq \sum\limits_{n=1}^{n_{cr}} n \, P(n) 
\end{equation}
for such a metastable state.
In this case from~(\ref{eq:c}) we obtain
\begin{equation}
\frac{c}{1-c} \simeq \frac{\sum\limits_{n=1}^{n_{cr}}
n \, \langle w \rangle^n \prod\limits_{m=1}^n \displaystyle{\frac{1}{w_m}} }
{1 + \sum\limits_{n=1}^{n_{cr}} 
\langle w \rangle^n \prod\limits_{m=1}^n \displaystyle{\frac{1}{w_m}} } \;.
\label{eq:met}
\end{equation}
instead of~(\ref{eq:uuu}) for calculation of $\langle w \rangle$ in 
this homogeneous metastable state. The critical cluster size is found self consistently
solving~(\ref{eq:k_cr}) and~(\ref{eq:met}) as a system of equations.
From~(\ref{eq:k_cr}) we see that the critical cluster size $n_{cr}$ 
diverges at $c \to c_{cr}$, 
since $\langle w \rangle \to w_{\infty}$. The results of calculation of $n_{cr}$
(with rounding down to an integer value) at $\sigma=0.5$, $b=1$
and at $\sigma=1$, $b=3$  are shown in 
Figs.~\ref{kcr_sig} and~\ref{kcr_b}, respectively.

\section{The mean nucleation time}

To our knowledge, the nucleation time, i.~e., the mean time the system 
stays in the homogeneous metastable state, has not been 
considered up to now for the 
zero--range process. A characteristic time scale for a particle to 
escape from a jam has been evaluated in~\cite{LEC98} for the 
related bus route
model. This time scale, however, refers to already formed finite--size 
jams and is therefore different to the nucleation time we are interested in.

Within the framework of mean-field dynamics, the mean nucleation time in
our model can be evaluated as follows. 
Let ${\mathcal P}(t)$ be the probability density of the first passage 
time to exceed the critical number of particles $n_{cr}$ in a single box. 
By our definition, the nucleation occurs when one of the $M$ boxes reaches the 
cluster size $n_{cr}+1$. The probability that it occurs first in a given
box within a small time interval $[t;t+dt]$ is thus 
${\mathcal P}(t) dt \times \left[1 - \int\limits_0^t {\mathcal P}(t') dt' \right]^{M-1}$
according to our assumption that the boxes are statistically independent.
The term $\left[1 - \int\limits_0^t {\mathcal P}(t') dt' \right]^{M-1}$ 
is the probability that in all other boxes, except
the given one, the overcritical cluster size $n_{cr}+1$ has still not been reached.
Since, the nucleation can occur in any of $M$ boxes, the nucleation
probability density ${\mathcal P}_M(t)$ for the system of $M$ boxes is given by
\begin{eqnarray}
{\mathcal P}_M(t) &=& M \, {\mathcal P}(t) \times 
\left[1 - \int\limits_0^t {\mathcal P}(t') dt' \right]^{M-1} \nonumber \\
&\simeq& M {\mathcal P}(t) \exp \left( -M \int\limits_0^t {\mathcal 
P}(t') dt' \right) \;.
\label{eq:PM}
\end{eqnarray}
The latter equality holds for large $M$, since all $M$ boxes are equivalent, and 
therefore the probability $\int\limits_0^t {\mathcal P}(t') dt'$ that the nucleation 
occurs in a given box within a characteristic time interval $t \sim 
\langle T \rangle_M$ is a small quantity of order $1/M$.
The mean nucleation time for the system of $M$ boxes is
\begin{equation}
\langle T \rangle_M = \int\limits_0^{\infty} t \, {\mathcal P}_M(t) dt \;.
\label{eq:TavM}
\end{equation}
Here $\langle T \rangle_1$ is the mean first passage time
for a single box. 

In order to estimate $\langle T \rangle_M$ according to~(\ref{eq:PM}) 
and~(\ref{eq:TavM}), 
one needs some idea about the first passage time probability density
for one box ${\mathcal P}(t)$. 
This is actually the problem of a particle escaping from a potential well.
Since we start with an almost homogeneous state of the system, 
we may assume zero cluster size $n=0$ as the initial condition. 
The first passage time probability density can be calculated
as the probability per unit
time to reach the state $n_{cr}+1$, assuming that the
particle is absorbed there. It is reasonable to assume that after
a certain equilibration time $t_{eq}$, when a 
quasi--stationary distribution
of the cluster sizes within $n \le n_{cr}$ is reached, the escaping from this region  
is characterised by a certain transition rate $w_{esc}$.
Hence, for $t>t_{eq}$ we have
\begin{equation}
{\mathcal P}(t) \simeq w_{esc} \times \left[ 1 - \int\limits_0^t 
{\mathcal P}(t') dt' \right] \;,
\label{eq:Pint}
\end{equation}
where the expression in square brackets is the probability that 
the absorption at $n_{cr}+1$ has still not occured up to the
time moment $t$.
At high enough potential barriers (large mean first passage times)
the short--time contribution to the integral is irrelevant and, by 
means of~(\ref{eq:TavM}), the solution of~(\ref{eq:Pint}) can be written as
\begin{equation}
{\mathcal P}(t) =  \frac{1}{\langle T \rangle_1} \exp
\left( -\frac{t}{\langle T \rangle_1} \right) \;,
\label{eq:sol1}
\end{equation}
where $\langle T \rangle_1 = w_{esc}^{-1}$.
Obviously, this approximate solution of the first passage problem
is not valid for very short times $t \ll t_{eq}$, since the short--time solution should
explicitly depend on the initial condition. In particular, if we start at $n=0$,
then the state $n_{cr}+1$ cannot be reached immediately, so that ${\mathcal P}(0)=0$.
Nevertheless~(\ref{eq:sol1}) can be used to estimate the mean nucleation time
$\langle T \rangle_M$ provided that $\langle T \rangle_M > t_{eq}$. 

\begin{figure}
\begin{center}
\includegraphics[scale=0.3]{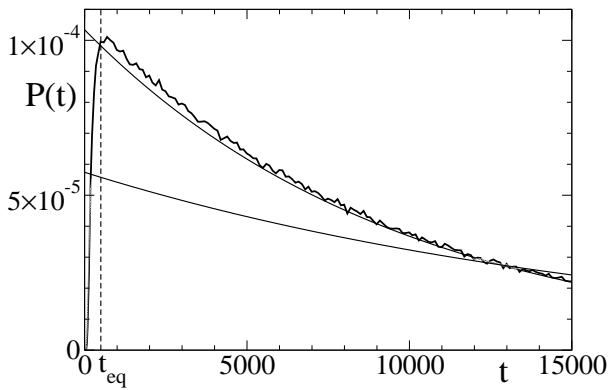}     
\end{center}
\hspace{8mm}
\begin{center}
\includegraphics[scale=0.3]{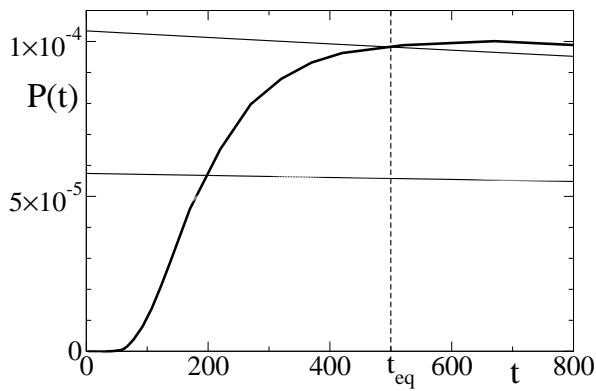}	
\end{center}
\caption{Comparison between the theoretical 
approximation~(\ref{eq:sol1}) for ${\mathcal P}(t)$ (smooth curves) and 
mean--field simulation results
(fluctuating curves) shown in a longer (top) and in a shorter (bottom) 
time scale. The vertical dashed line 
indicates the equilibration time $t_{eq} \approx 500$ for the set of parameters 
$\sigma=0.5$, $b=1$, and $c=0.84$ represented by the upper curves in 
both pictures. The other curves correspond to $\sigma=1$, $b=3$, and $c=0.61$. 
}
\label{fptd}
\end{figure}

We have checked the correctness of these theoretical expectations 
within the mean--field dynamics represented by~(\ref{eq:master}) 
and~(\ref{eq:master0})
by comparing them with the results of simulation of stochastic
trajectories generated according to these equations.
The simulation curves for ${\mathcal P}(t)$ at two 
different sets of parameters: $\sigma=0.5$, $b=1$, $c=0.84$ (with $n_{cr}=48$), 
and $\sigma=1$, $b=3$, $c=0.61$ (with $n_{cr}=35$) are shown in 
Fig.~\ref{fptd} in two different time scales. 
As we see, Eq.~(\ref{eq:sol1}) is a good 
approximation for large enough times $t>t_{eq}$. For definiteness, we have identified 
the equilibration time $t_{eq}$ with the 
crossing point of the theoretical and simulated curves.
An interesting additional feature is the presence of an apparent 
nucleation time lag, which is about $t_{lag} \approx 60$ for the first set of 
parameters and about $t_{lag} \approx 30$ for the second one. 
Evidently, the first passage time probability density ${\mathcal P}(t)$ 
tends to zero very rapidly when $t$ decreases below $t_{lag}$.

By inserting~(\ref{eq:sol1}) in~(\ref{eq:TavM}) we obtain
\begin{equation}
\langle T \rangle_M \simeq M \, \langle T \rangle_1 
\int\limits_0^{\infty} x \, e^{-x} \exp \left( -M \left[ 1 - e^{-x} \right] \right) dx 
\end{equation}
after changing the integration variable $t / \langle T \rangle_1 \to x$.
Taking into account that only the region $x \sim 1/M$ contributes to the integral
at large $M$, 
we arrive at a very simple expression
\begin{equation}
\langle T \rangle_M \simeq \frac{ \langle T \rangle_1}{M} 
\label{eq:simple}
\end{equation}
relating the mean first passage time or nucleation time in a system of
$M$ boxes with that of one box. 
The latter can be calculated easily by the known formula~\cite{Gardiner}
\begin{equation}
\langle T \rangle_1 = \sum\limits_{n=0}^{n_{cr}} 
\left[ \langle w \rangle \tilde P(n) \right]^{-1} \sum\limits_{m=0}^n \tilde P(m) \;,
\label{eq:mfpt}
\end{equation}
where $\tilde P(0) = 1$ and $\tilde P(n)= \prod\limits_{k=1}^{n} ( \langle w \rangle / w_k)$
with $n>1$ represent the unnormalised stationary probability distribution.

\begin{figure}
\begin{center}
\includegraphics[scale=0.3]{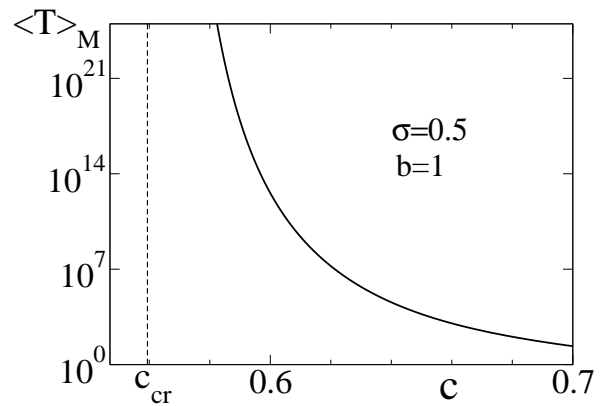}
\end{center}
\caption{Mean nucleation time vs density at $\sigma=0.5$, $b=1$,
and $M=10^6$.
The critical density is indicated by a vertical dashed line.}
\label{nuclt_sig}
\end{figure}

\begin{figure}
\begin{center}
\includegraphics[scale=0.3]{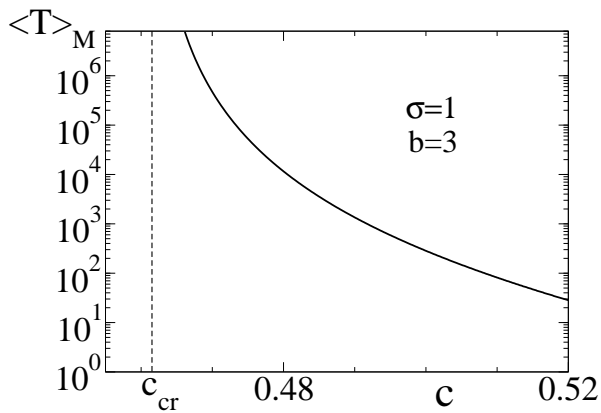}
\end{center}
\caption{Mean nucleation time vs density at $\sigma=1$, $b=3$,
and $M=10^6$.
The critical density is indicated by a vertical dashed line.}
\label{nuclt_b}
\end{figure}

The mean nucleation time versus
density $c$, calculated from~(\ref{eq:simple}) and~(\ref{eq:mfpt})
at $M=10^6$ is shown in Figs.~\ref{nuclt_sig} and~\ref{nuclt_b}. 
Fig.~\ref{nuclt_sig} refers to the
case $\sigma=0.5$ and $b=1$, whereas Fig.~\ref{nuclt_b} 
is for the case $\sigma=1$ and $b=3$. 
These figures show that the mean nucleation time increases dramatically as the 
critical cluster size $n_{cr}$ increases (see the corresponding plots in 
Figs.~\ref{kcr_sig} and~\ref{kcr_b}) approaching the critical density $c_{cr}$.

According to our previous discussion,
estimate~(\ref{eq:simple}) is valid for large enough mean nucleation times
$\langle T \rangle_M > t_{eq}$, in particular, when approaching the
critical density $c \searrow c_{cr}$ at any large but fixed $M$.
It is not valid in the thermodynamic limit $M \to \infty$ at a fixed density $c$
---Eq.~(\ref{eq:simple}) suggests that $\langle T \rangle_M$ decreases 
as $\sim 1/M$, whereas in reality the decrease must be 
slower for small nucleation times (large $M$) since ${\mathcal P}(t) \to 0$ as $t \to 0$.
In particular, the mean--field dynamics suggests that for a wide range of $M$ values
$\langle T \rangle_M$ quasi--saturates
at $\langle T \rangle_M \approx t_{lag}$, since the critical 
cluster size is almost never reached before $t=t_{lag}$.

\section{Simulation results}
\label{s:sim}

\begin{figure}
\begin{center}
\includegraphics[scale=1.1]{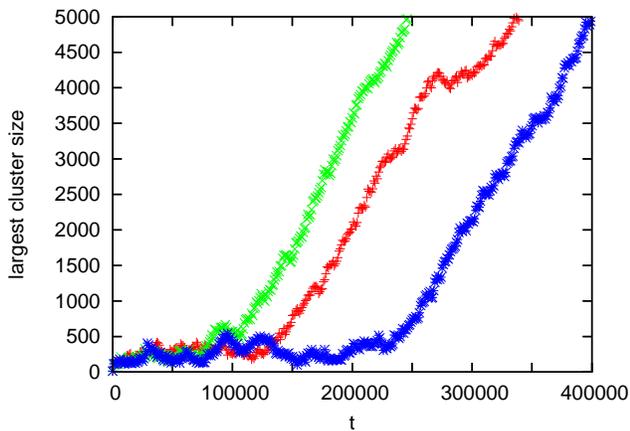}
\end{center}
\caption{(Color online). 
Largest cluster size versus time for $\sigma=0.5$, $b=1$, $w_1=5$, $c=0.66$, $M=10^5$. 
Results from three independent Monte Carlo runs are shown.}
\label{sim}
\end{figure}

Numerical simulations of the zero-range model show clear evidence for
the existence of a metastable state prior to condensation.  
In Fig.~\ref{sim} we show the largest cluster size as a function of time for three 
separate Monte Carlo runs in the case $\sigma=0.5$, $b=1$, $w_1=5$, $M=10^5$.
For each run the system was started in a random uniform initial
condition with density $c=0.66$ (for these parameters $c_{cr} \simeq
0.56$).  It is clearly seen that after a short equilibriation period the
system fluctuates in a metastable state before a condensate appears. 
The critical cluster size is observed to be around 400 in good
agreement with the prediction 
 $n_{cr} \simeq 330$ from
Eqs.~\eqref{eq:k_cr} and~\eqref{eq:met} (see Fig.~\ref{kcr_sig}).
However, the metastable time is about an order of magnitude larger than
predicted.

\begin{figure}
\begin{center}
\includegraphics[scale=1.1]{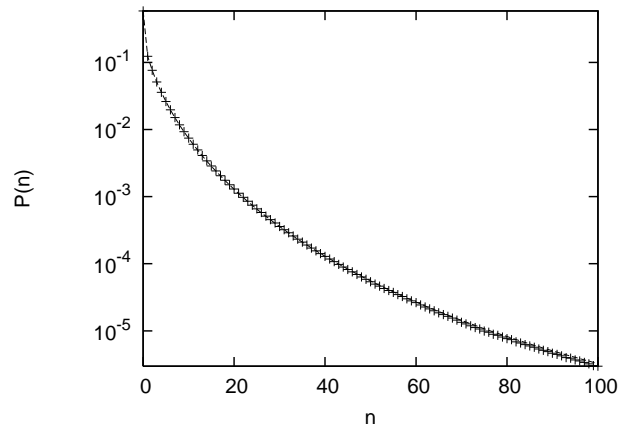}  
\end{center}
\caption{Distribution over small cluster sizes in metastable state: 
results for Monte Carlo simulation (crosses) compared to prediction 
of~\eqref{eq:Pk} with $\langle w \rangle = w_{n_{cr}}$ and $n_{cr}$ 
calculated numerically from~\eqref{eq:k_cr} and~\eqref{eq:met}
(dashed line).}. 
\label{dist}
\end{figure}

In Fig.~\ref{dist} we show the distribution of cluster sizes (for small
clusters) averaged over the metastable state of one such run.  The
distribution is in good agreement with 
Eq.~\eqref{eq:Pk} with $\langle w \rangle = w_{n_{cr}}$, 
thus supporting the assumption of quasi--stationarity.

In the analytical treatment of the previous section we
calculated the mean time for the maximum cluster size to exceed $n_{cr}$
under the assumption that the
current in the metastable state is constant.  In practice, of course,
the metastable current also fluctuates (and the fluctations are greater
when $w_n$ depends more strongly on cluster size, e.~g., for the
$\sigma=1$ case compared to, say, $\sigma=0.5$).  Simulations suggest
that these fluctuations can destroy the metastable state in cases where
the metastable current $w_{n_{cr}}$ is close to the
current of the condensed phase $w_\infty$.

\begin{figure}
\begin{center}
\includegraphics[scale=1.1]{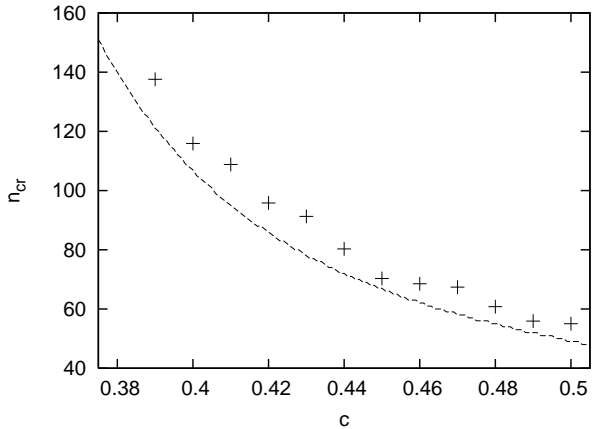}  
\end{center}
\caption{Critical cluster size versus density for 
$\sigma=0.5$, $b=3$, $w_1=5$, $M=10^5$ ($c_{cr} \simeq 0.27$).  
Crosses show simulation data (averaged over 10 Monte Carlo histories), 
dashed line is prediction of Eqs.~\eqref{eq:k_cr} and~\eqref{eq:met}.}. 
\label{densclust}
\end{figure}

\begin{figure}
\begin{center}
\includegraphics[scale=1.1]{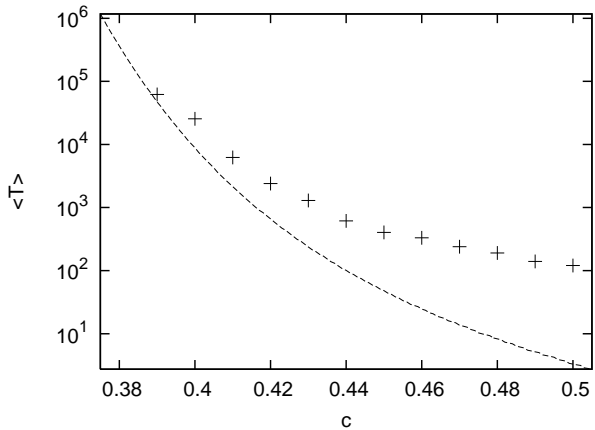}  
\end{center}
\caption{Nucleation time versus density for 
$\sigma=0.5$, $b=3$, $w_1=5$, $M=10^5$ ($c_{cr} \simeq 0.27$).  
Crosses show simulation data (averaged over 10 Monte Carlo histories), 
dashed line is prediction of Eqs.~\eqref{eq:simple} and~\eqref{eq:mfpt}.}.  
\label{denstime}
\end{figure}

In contrast, for parameters where the metastable state is well-separated
from the condensed state we find relatively good quantitive agreement
between theory and simulation.  For example, in Figs.~\ref{densclust}
and~\ref{denstime} we compare the average simulation values of critical
cluster size and nucleation time to the theoretical predictions for a
range of densities in the case $\sigma=0.5$ and $b=3$.
In these simulations we crudely identified the end of the metastable
state as the point when the current out of the largest cluster had been
less than the average system current for 50 consecutive Monte Carlo time
steps. 

We find that our mean-field theory fairly accurately reproduces the
critical cluster size but systematically underestimates the nucleation
time. This discrepancy may be partly
due to the presence of (weak) dynamical correlations between
the numbers of particles in the boxes in the
fluctuating metastable state. Namely, 
the appearance of a large cluster with $n \simeq n_{cr}$ is 
likely to be accompanied by a slight depletion of the surrounding 
medium.  
Furthermore, we only calculated the mean first passage time and ignored
the probability  that a cluster reaches $n_{cr}+1$ and is immediately
driven by a fluctuation back below $n_{cr}$.   Monte Carlo histories
which involve such a fluctuation back into the metastable state before a
condensate is established, would increase the average simulation
nucleation time above the theoretical prediction.

Despite the neglect of current fluctuations, dynamical correlations,
etc., our simulations show that
the simple mean-field approach provides a good qualitative description
of the metastable state and its dependence on density. It thus
represents an important first step towards more refined theories.

\begin{figure}
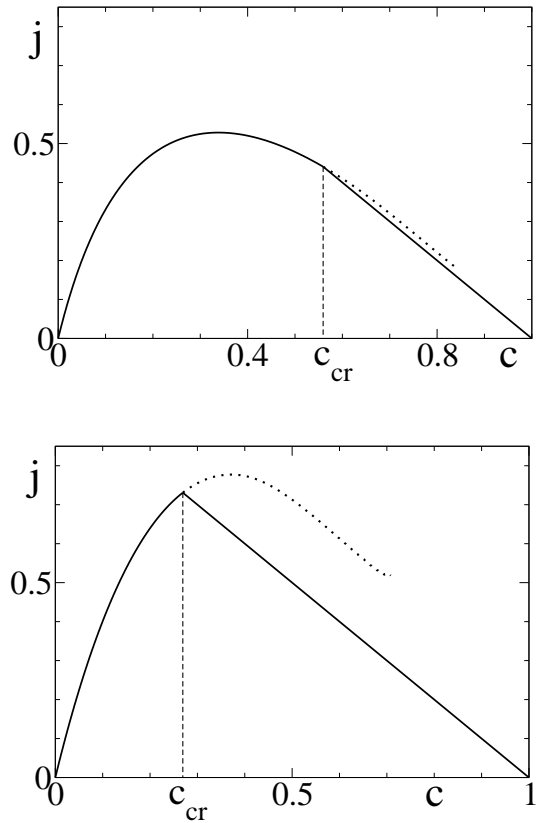

\begin{center}
\includegraphics[scale=0.3]{zrtraff12a}\\ 
\vspace*{8mm}
\includegraphics[scale=0.3]{zrtraff12b}\\ 
\end{center}
\caption{The fundamental (flux--density) diagram for
two different sets of control parameters:
$\sigma=0.5$, $b=1$, $w_1=5$ (top); $\sigma=0.5$, $b=3$, $w_1=5$ 
(bottom).
The branches
of metastable homogeneous state are shown by dotted lines,
the critical densities $c_{cr}$ are indicated by vertical dashed lines.}
\label{fund}
\end{figure}

\section{Conclusions}

In this paper we have studied a zero-range model motivated by the observation of phase 
separation in traffic flow. In particular, we have demonstrated the existence of a 
metastable state which exists prior to condensation and obtained analytical estimates 
for the critical cluster size and mean nucleation time.  This new insight into the mechanism 
of cluster formation in the ZRP is supported by the simulation results discussed in the 
previous section. In the context of traffic flow we now conclude by further exploring 
the significance of this metastable state.

The relation between density $c$ and flux $j$ of cars is known as the fundamental
diagram of traffic flow. The average stationary flux can be calculated as follows
\begin{equation}
j = \sum\limits_{n=1}^{\infty} Q(n) \, w_n \;,
\label{eq:j0}
\end{equation}
where $Q(n)$ is the probability that there is a car in a given cell
(in the original model) which can move forwards with the rate $w_n$.
Note that only those cars contribute to the flux, which are the first in
some cluster. Hence, $Q(n) = \varphi P(n) / \sum\limits_{m=1}^{\infty} P(m)$,
where $\varphi$ is the fraction of cells, which contain such cars. 
This fraction can be calculated easily as the number of clusters divided by the 
total number of cells. These quantities
fluctuate in our model. For large systems, however, they can be
replaced by the mean values. The mean number of clusters is equal to the
mean number of non-empty boxes $M \sum\limits_{n=1}^{\infty} P(n)$ in the
zero-range model,
whereas the mean number of cells, i.~e., the mean length of the road is
$\langle L \rangle = M + \langle N \rangle = M \, (1+ \langle n \rangle)
= M/(1-c)$, as we have already discussed in Sec.~\ref{sec:master}.  
Hence, $Q(n) = (1-c) \, P(n)$ and~(\ref{eq:j0}) reduces to
\begin{equation}
j = (1-c) \, \langle w \rangle \;.
\label{eq:j}
\end{equation}
The mean stationary transition rate $\langle w \rangle$ depends on the car 
density $c$. For undercritical densities $c<c_{cr}$, this quantity is the
solution of~(\ref{eq:uuu}). For overcritical densities we have 
$\langle w \rangle = w_{\infty}$ in the phase coexistence regime,
as discussed in Sec.~\ref{sec:trr}, therefore in this case the
fundamental diagram reduces to a straight line
\begin{equation}
j = (1-c) \, w_{\infty} \quad : \quad c \ge c_{cr} \;.
\end{equation}
In the metastable homogeneous state at $c>c_{cr}$ the mean transition
rate $\langle w \rangle$ together with the critical cluster size $n_{cr}$
can be found from the system of equations~(\ref{eq:k_cr}) and~(\ref{eq:met}),
which allows calculation of the metastable branch of flux $j$.

The resulting fundamental diagrams for $\sigma=0.5$, $w_1=5$ and two values of 
parameter $b$ are shown in Fig.~\ref{fund}. As we see, the shape of the
fundamental diagram, as well as the critical density and location of the 
metastable branch depend remarkably on the value of $b$. 
These features will also depend on the values of $\sigma$ and $w_1$. 
The metastable branch ends abruptly at certain density above which
Eqs.~(\ref{eq:k_cr}) and~(\ref{eq:met}) have no real solution.
It corresponds to a relatively small, but finite value of the critical
cluster size $n_{cr}$.

Note that a metastable branch is also observed in simulations of
cellular automata with slow--to--start 
rules~\cite{Takayasu93,Benjamin96,Barl98}. In our examples, however, 
the metastable
 branch is located at larger densities and decreases with
increasing 
 of $c$ over a certain wide range of values depending on
$b$, $\sigma$ and $w_1$. The simulations of the previous
section suggest that when this metastable branch is well separated from
the condensed section of the fundamental diagram, our picture is robust
even in the presence of fluctuations.

In summary therefore, we believe that by suitable variation of parameters our 
simple model can reproduce some important features of real traffic flow.  
There is much scope for further investigation, both analytical and computational.

\appendix*

\section{Critical cluster distribution for $0<\sigma<1$}

Here we focus on the cluster size distribution at $c=c_{cr}$ in
the case $0<\sigma<1$, presenting a detailed calculation of the
asymptotic behaviour for large $n$.

According to~(\ref{eq:Pk}), 
at $c=c_{cr}$ where $\langle w \rangle = w_{\infty}$,
we have
\begin{eqnarray}
\label{eq:lnP}
&&\ln P(n) = \ln \left( \frac{P(0) w_{\infty}}{w_1} \right) 
- \sum\limits_{m=2}^n \ln \left( 1 + \frac{b}{m^{\sigma}} \right) \\
&=& \ln \left( \frac{P(0) w_{\infty}}{w_1} \right)
- \int\limits_1^n \ln \left( 1 + \frac{b}{m^{\sigma}} \right) \, dm
+ C + \delta(n) \nonumber \;,
\end{eqnarray}
where $C$ is a constant, and $\delta(n) \to 0$ at $n \to \infty$.
The latter follows from the fact that each term with $m=k$ in the sum
generates terms $\propto k^{-y}$ with $y>0$ when the logarithm is 
expanded in Taylor series, and for each of these terms we have
$k^{-y}= \int\limits_{k-1}^k m^{-y} dm + O \left( k^{-y-1} \right)$
at $k \to \infty$. It ensures that the difference between the integral 
and the sum in~(\ref{eq:lnP}) is finite and tends to some constant 
$C$ at $n \to \infty$ for $\sigma>0$. The remainder term $\delta(n)$ is 
irrelevant for the leading asymptotic behaviour of 
$P(n)$. By expanding the logarithm and integrating term by term, 
for $0 < \sigma <1$ and $\sigma \ne 
\frac{1}{2},\frac{1}{3},\frac{1}{4},\ldots$ we obtain 
\begin{equation}
P(n) \propto \prod\limits_{k=1}^{[1/\sigma]} 
\exp \left\{ \frac{(-b)^k}{k} \frac{n^{1-k\sigma}}{1-k\sigma} \right\} \;,
\label{eq:Pas1}
\end{equation}
where $[1/\sigma]$ denotes the integer part of $1/\sigma$.
The cases where $\sigma$ is an inverse integer are
special, since a term $\propto 1/m$ appears in the expansion of the 
logarithm, giving rise to a power--like correction
to the stretched exponential behaviour, viz.
\begin{equation}
P(n) \propto n^{\sigma (-b)^{1/\sigma}} 
\prod\limits_{k=1}^{\sigma^{-1}-1} 
\exp \left\{ \frac{(-b)^k}{k} \frac{n^{1-k\sigma}}{1-k\sigma} \right\}
\label{eq:Pas2}
\end{equation}
for $\sigma = \frac{1}{2},\frac{1}{3},\frac{1}{4},\ldots$.
The known result for $\sigma=1$ can also be obtained by this method:
it corresponds to the power--like prefactor in~(\ref{eq:Pas2}).
Only the linear expansion term of the logarithm is relevant at 
$1/2 < \sigma<1$, so we find
$P(n) \propto \exp \left[ -b \, n^{1-\sigma}/(1-\sigma) \right]$ 
in the limit of large $n$.
The first two terms are relevant for $1/3 < \sigma \le 1/2$,
the third one becomes important for $1/4 < \sigma \le 1/3$, and so on.
Eqs.~(\ref{eq:Pas1}) and~(\ref{eq:Pas2}) represent an exact analytical 
result at $n \to \infty$
which we have also verified
numerically at different values of 
$\sigma$ and $b$. In this form, where the proportionality coefficient 
is not specified, Eqs.~(\ref{eq:Pas1}) and~(\ref{eq:Pas2}) are 
universal, i.~e., they do not depend on the choice of $w_1$.

\begin{acknowledgments}
One of us (J.~K.) acknowledges support by the
German Science Foundation
(DFG) via project No. 436 LET 17/3/05. The main part of the research
was done at Rostock University, Institute of Physics.
We are grateful to G.~M.~Sch\"utz (J\"ulich) for helpful discussions 
and comments on the manuscript.
\end{acknowledgments}


\begin{thebibliography}{}

\bibitem{Mukamel00}
D.~Mukamel, Phase transitions in nonequilibrium systems, in 
{\em Soft and Fragile Matter: Nonequilibrium Dynamics,
  Metastability and Flow}, edited by M.~E. Cates and M.~R. Evans, Institute of
  Physics Publishing, Bristol, 2000

\bibitem{Schutz03}
G.~M. Sch{\"u}tz, J. Phys. A: Math. Gen. {\bf 36}, R339 (2003)

\bibitem{S1} Y. Kafri, E. Levine, D. Mukamel, G. M. Sch\"utz, J. T\"or\"ok,
Phys. Rev. Lett. \textbf{89}, 035702 (2002)

\bibitem{Evans05}
M.~R. Evans, T.~Hanney,
J. Phys. A: Math. Gen. \textbf{38}, R195 (2005)

\bibitem{physrep} R. Mahnke, J. Kaupu\v{z}s, I. Lubashevsky,
Physics Reports \textbf{408}, Issue 1--2, pp. 1--130 (2005)

\bibitem{S3} S. Grosskinsky, G. M. Sch\"utz, H. Spohn,
J. Stat. Phys. \textbf{113}, 389 (2003)

\bibitem{Godreche03}
C. {Godr{\`e}che}, J. Phys. A {\bf 36}, 6313 (2003)

\bibitem{cssrev} D. Chowdhury, L. Santen, A. Schadschneider, 
Physics Reports \textbf{329}, 199 (2000)

\bibitem{Helbing01}
D.~Helbing, Rev. Mod. Phys. {\bf 73}, 1067 (2001)

\bibitem{mk99} R. Mahnke, J. Kaupu\v{z}s, Phys. Rev. E \textbf{59}, 117 (1999)

\bibitem{mkf03} R. Mahnke, J. Kaupu\v{z}s, V. Frishfelds, Atmos. Res. \textbf{65}, 261 (2003)

\bibitem{Levine04}
E. Levine, G. Ziv, L. Gray, D. Mukamel, Physica A {\bf 340}, 636 (2004)

\bibitem{Spitzer70}
F.~Spitzer, Adv. Math. {\bf 5}, 246 (1970)

\bibitem{Evans00}
M.~R. Evans, Braz. J. Phys. {\bf 30}, 42 (2000)

\bibitem{LEC98} O' Loan, M. R. Evans, M. E. Cates, 
Phys. Rev. E {\bf 58}, 1404 (1998)

\bibitem{Takayasu93}
M.~Takayasu, H.~Takayasu, Fractals {\bf 1}, 860 (1993)

\bibitem{Benjamin96}
S.~C. Benjamin, N.~F. Johnson, P.~M. Hui, J. Phys. A {\bf 29}, 3119 (1996)

\bibitem{Barl98}
R.~Barlovic, L.~Santen, A.~Schadschneider, M. Schreckenberg, 
Eur. Phys. J. B {\bf 5}, 793 (1998)

\bibitem{km00} J. Kaupu\v{z}s, R. Mahnke, Eur. Phys. J. B \textbf{14}, 793 (2000)

\bibitem{S2} Y. Kafri, E. Levine, D. Mukamel, G. M. Sch\"utz, R. D. Willmann,
Phys. Rev. E \textbf{68}, 035101(R) (2003)

\bibitem{Jeon00}
I.~Jeon, P.~March, Can. Math. Soc. Conf. Proc. {\bf 26}, 233 (2000)

\bibitem{GSprivate} G. M. Sch\"utz, private communication (2005)

\bibitem{Gardiner} C. W. Gardiner, \textit{Handbook of Stochastic Methods for
Physics, Chemistry, and the Natural Sciences}, Springer, Berlin, 1983, 1994

\end{thebibliography}
\end{document}